\documentclass[preprint,showpacs,preprintnumbers,amsmath,amssymb]{revtex4}

\usepackage{graphicx}
\usepackage{dcolumn}
\usepackage{bm}

\begin{document}

\preprint{}

\title{Advances in delimiting the Hilbert-Schmidt separability probability of real two-qubit systems}
\author{Paul B. Slater}%
\email{slater@kitp.ucsb.edu}
\affiliation{%
ISBER, University of California, Santa Barbara, CA 93106\\
}%
\date{\today}

\begin{abstract}
We seek to derive the probability--expressed in terms of the Hilbert-Schmidt (Euclidean or flat) metric--that a generic (nine-dimensional) real two-qubit system is separable, by implementing the well-known Peres-Horodecki test on the partial transposes (PT's) of the associated $4 \times 4$ density matrices ($\rho$). But the full implementation of the test--requiring that the determinant of the PT be nonnegative for separability to 
hold--appears to be, at least presently, computationally intractable.
So, we have previously implemented--using the auxiliary concept of a {\it diagonal-entry-parameterized separability function} (DESF)--the weaker implied test of nonnegativity of the six $2 \times 2$ principal minors of the PT. This yielded an exact 
upper bound on the separability probability of 
$\frac{1024}{135 \pi^2} \approx 0.76854$. Here, we piece together (reflection-symmetric) results obtained by requiring that each of the four $3 \times 3$ principal minors of the PT, in turn, be nonnegative, giving an improved/reduced upper bound of 
$\frac{22}{35} \approx 0.628571$. Then, we conclude that a still 
further improved upper bound of 
$\frac{1129}{2100} \approx 0.537619$ can be found by similarly piecing together the (reflection-symmetric) results of enforcing the {\it simultaneous} nonnegativity of 
certain pairs of the four $3 \times 3$ principal minors.
Numerical simulations--as opposed to exact symbolic 
calculations--indicate, on the other hand, that the true probability is certainly less than 
$\frac{1}{2}$. Our analyses lead us to suggest a possible form for the true DESF, yielding a separability probability of 
$\frac{29}{64} \approx 0.453125$, while the {\it absolute} separability probability of 
$\frac{6928-2205 \pi }{2^{9/2}} \approx 0.0348338$ provides the best exact lower bound established so far. In deriving our improved upper bounds, we rely repeatedly upon the use of certain
integrals over cubes that arise. Finally, we apply an independence assumption to a pair of DESF's  that comes close to reproducing our numerical estimate of the true separability function.
\end{abstract}


\maketitle

{\.Z}yczkowski, Horodecki, Sanpera and Lewenstein, in a much-cited article \cite{ZHSL},  have given "philosophical", "practical" and "physical" reasons for studying "separability probabilities". We have examined the associated problems which arise, using the volume elements of several metrics of interest as measures on the quantum states, in various numerical and theoretical studies \cite{slaterA,slaterC,slaterJGP,slaterPRA,pbsCanosa,slater833,JMP2008,ratios}. 

\begin{figure}
\includegraphics{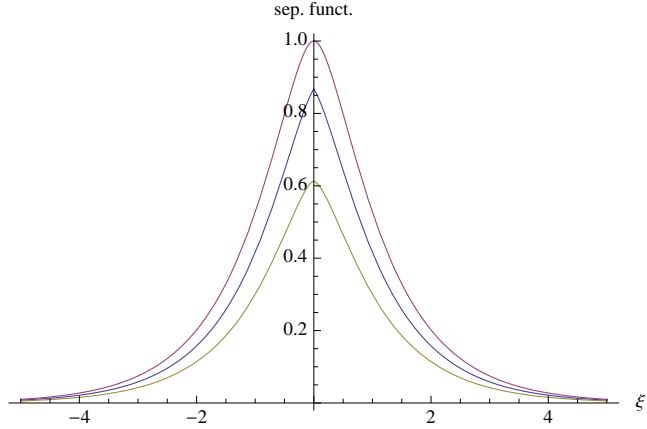}
\caption{\label{fig:threecurves}Three forms of diagonal-entry-parameterized separability functions (DESF's)}
\end{figure}

In these regards, we begin our presentation by directing the reader's attention to Fig.~\ref{fig:threecurves}. These depicts various forms of "diagonal-entry-parameterized separability functions" (DESF's) \cite{slaterPRA2,slater833}--as opposed to "eigenvalue-parameterized separability functions (ESF's) \cite{maxconcur4,maxconcur2,JMP2008}--that we will employ here to obtain estimates and simple exact upper bounds on the Hilbert-Schmidt (HS) probability that a generic (nine-dimensional) real two-qubit system is separable.

\begin{figure}
\includegraphics{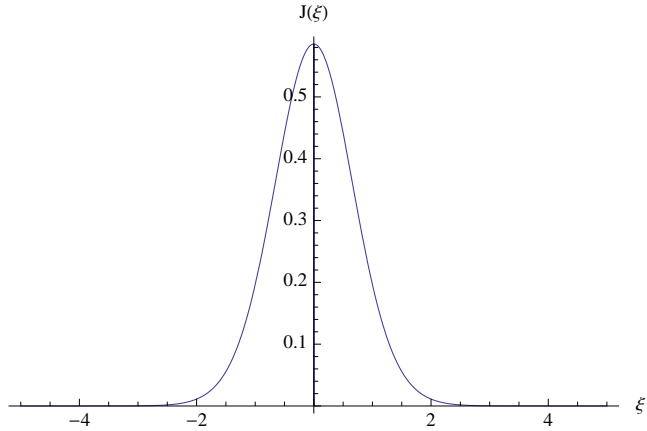}
\caption{\label{fig:JacobianMinors}Jacobian (\ref{jacobian}), which when multiplied by a separability function and integrated over $\xi \in [-\infty,\infty]$ yields the associated Hilbert-Schmidt separability probability}
\end{figure}

The subordinate of the three curves in 
Fig.~\ref{fig:threecurves}--derived using an extensive quasi-Monte Carlo (Tezuka-Faure \cite{giray1,tezuka}) six-dimensional numerical integration procedure--provides an estimate of the true, but so-far not exactly-determined DESF. The dominant of the three curves--readily obtainable from results reported in 
\cite[sec. VII]{slaterPRA2}--has the form
\begin{equation} \label{dominantcurve}
S_{dom}(\xi)=\begin{cases}
\frac{1}{2} e^{-3 \xi } \left(3 e^{2 \xi }-1\right) & \xi >0 \\
 -\frac{1}{2} e^{\xi } \left(e^{2 \xi }-3\right) & \xi <0
\end{cases}.
\end{equation}
The intermediate of the three curves, which we first report here, has the same--differing only in constants--functional form
\begin{equation} \label{intermediatecurve}
S_{int}(\xi)= \begin{cases}
 \frac{9 \pi^2}{2048}  e^{-3 \xi } \left(27 e^{2 \xi }-7\right) & \xi >0
   \\
 -\frac{9 \pi^2}{2048}  e^{\xi } \left(7 e^{2 \xi }-27\right) & \xi <0
\end{cases}.
\end{equation}
With each of these three curves we can obtain an associated estimate or upper bound on
the desired HS separability probability ($P^{HS}_{sep/real}$). This is accomplished by integrating over $\xi \in [-\infty,\infty]$ the product of the corresponding curve with the function (Fig.~\ref{fig:JacobianMinors}) (based on the jacobian of a 
coordinate transformation, to be described below)
\begin{equation} \label{jacobian}
J(\xi)= \frac{64 \text{csch}^9(\xi ) (-160 \sinh (2 \xi )-25 \sinh (4 \xi )+12
   \xi  (16 \cosh (2 \xi )+\cosh (4 \xi )+18))}{27 \pi ^2},
\end{equation}
that is,
\begin{equation} \label{JACOBIAN}
P^{HS}_{sep/real}=\int^{\infty}_{-\infty} S(\xi) J(\xi) \mbox{d} \xi.
\end{equation}
Proceeding, thusly, we obtain an upper bound on the HS separability probability of 
$\frac{1024}{135 \pi^2} \approx 0.76854$ based on the dominant of the three curves,    
$\frac{22}{35} \approx 0.628571$ using the intermediate curve, and an estimate of 0.4528427 for the true probability with the subordinate, numerically-derived curve. (From our work in \cite[eq. (25)]{maxconcur2}, we already know that the HS probability of a generic real two-qubit system being {\it absolutely} separable--that is not entanglable by any unitary transformation--is $\frac{6928-2205 \pi }{2^{9/2}} 
\approx 0.0348338$, which then serves as a {\it lower} bound on the corresponding HS [absolute {\it plus} nonabsolute] separability probability itself (cf. \cite{sepsize1} \cite[eq. (29)]{maxconcur2}).)

The variable $\xi$ used in the above presentation is the logarithm of the square root of the 
ratio of
the product of the 11- and 44-entry of the associated real $4 \times 4$ density matrix 
($\rho$) to the product of the 22- and 33-entries, that is 
\begin{equation} \label{xidefinition}
\xi =\log{\sqrt{\frac{\rho_{11} \rho_{44}}{\rho_{22} \rho_{33}}}}= \frac{1}{2} \log{\frac{\rho_{11} \rho_{44}}{\rho_{22} \rho_{33}}}.
\end{equation}
(In our previous studies \cite{slaterPRA2,slater833}, we have employed the alternative variables, 
$\nu =\frac{\rho_{11} \rho_{44}}{\rho_{22} \rho_{33}}$ and 
$\mu =\sqrt{\frac{\rho_{11} \rho_{44}}{\rho_{22} \rho_{33}}}$, but now switch to the [more symmetric] form (\ref{xidefinition}). Importantly, only the "cross-product ratio" of diagonal entries is needed in our parameterization to test for separability, and not the individual entries themselves.)
The jacobian (\ref{jacobian}) used in our calculations is obtained by the transformation
of one of the diagonal entries, say, $\rho_{33}$, to $\xi$ and integrating the Hilbert-Schmidt 
(Lebesgue) volume element (of course, $\rho_{44}=1-\rho_{11}-\rho_{22}-\rho_{33}$) \cite[p. 13646]{andai}
\begin{equation} \label{andaiformula}
\mbox{d} V_{HS}= (\rho_{11} \rho_{22} \rho_{33} \rho_{44})^{\frac{3 \beta}{2}} 
\mbox{d} \rho_{11} \mbox{d} \rho_{22} \mbox{d} \rho_{33}, \hspace{.25in} \beta=1
\end{equation}
over $\rho_{11}$ and $\rho_{22}$ and normalizing the result. (To obtain the corresponding HS volume elements for the {\it complex} $4 \times 4$ density matrices, one must employ--conforming to a pattern familiar from random matrix theory--$\beta=2$, and 
$\beta=4$ in the {\it quaternionic} case (cf. \cite{szHS}).)

\begin{figure}
\includegraphics{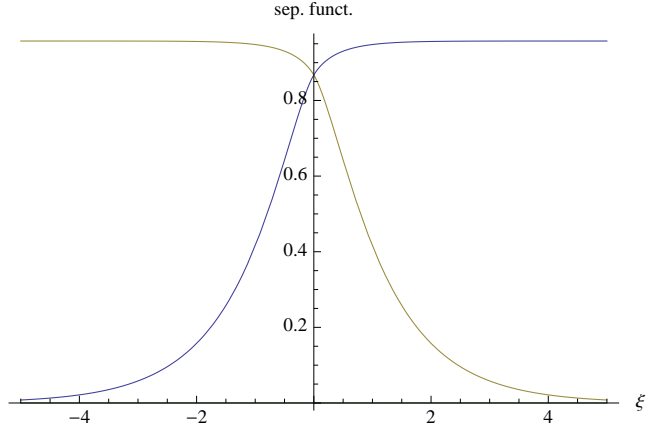}
\caption{\label{fig:SymmetricCurves}The two distinct (red and blue) separability functions obtained from the four $3 \times 3$ principal minors, the "envelope" (lesser branches) of which defines the intermediate curve in Fig.~\ref{fig:threecurves}. The $y$-intercepts of the two curves, are identically 
$\frac{45 \pi^2}{512} \approx 0.867446$.}
\end{figure}

\begin{figure}
\includegraphics{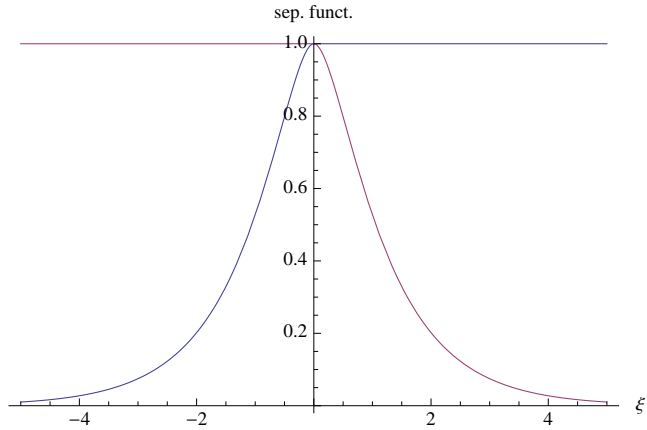}
\caption{\label{fig:SymmetricCurves2}The two distinct (red and blue) separability functions (discontinuous at $\xi=0$) obtained from the six $2 \times 2$ principal minors, the "envelope" (lesser branches) of which defines the dominant curve in Fig.~\ref{fig:threecurves}}
\end{figure}

The use of the celebrated Peres-Horodecki separability test \cite{asher,michal} is central to our analyses. Ideally, we would be able to require that the determinant of the partial transpose of 
$\rho$ be nonnegative to guarantee separability \cite{augusiak,ver}. However, this has so far proved to be too computationally demanding a (fourth-degree, high-dimensional) task for us to enforce (cf. \cite[eq. (7)]{slaterPRA2}). But, in
\cite{slater833}, we did succeed in implementing the weaker implied test that all the six $2 \times 2$ principal minors of the partial transpose of $\rho$ be nonnegative, giving us the dominant curve in Fig.~\ref{fig:threecurves}. (Actually, only two of the minors differ nontrivially from the analogous set of [nonnegative, of course] minors of 
$\rho$ itself.) To derive the sharper intermediate curve here, we extended this approach to the four $3 \times 3$ principal minors. Actually, we found that requiring each of four minors, in turn, to be nonnegative, yielded two pairs of identical results. Further, one of these results 
\begin{equation} \label{oneresult}
S_{3 \times 3}(\xi)= \begin{cases}
\frac{9 \pi ^2 e^{-3 \xi } \left(27 e^{2 \xi }-7\right)}{2048} & \xi >0
   \\
 \frac{3 \pi  e^{-3 \xi } \left(e^{\xi } \sqrt{1-e^{2 \xi }} \left(37
   e^{2 \xi }+2 e^{4 \xi }+21\right)+3 \left(27 e^{2 \xi }-7\right) \sin
   ^{-1}\left(e^{\xi }\right)\right)}{1024} & \xi <0
\end{cases}
\end{equation}
could be obtained from the other set by the transformation $\xi \rightarrow -\xi$. This curve (\ref{oneresult}) and its reflection around $\xi=0$ are shown in Fig.~\ref{fig:SymmetricCurves}. The intermediate curve (\ref{intermediatecurve}) in Fig.~\ref{fig:threecurves}, first reported here, was constructed by joining the sharper segments of these two curves over the two half-axes. 
A parallel strategy had been pursued with the $2 \times 2$ minors. The comparable results to (\ref{oneresult}) and 
Fig.~\ref{fig:SymmetricCurves} for the $2 \times 2$ minors investigation \cite{slater833}
are
\begin{equation}
 S_{2 \times 2}(\xi)=\begin{cases}
 e^{-2 \xi } (2 \sinh (\xi )+\cosh (\xi )) & \xi >0 \\
1 & \xi <0 
\end{cases}
\end{equation}
and Fig.~\ref{fig:SymmetricCurves2}.

For the intermediate curve in Fig.~\ref{fig:threecurves} we have the nontrivial $y$-axis intercept of $\frac{45 \pi^2}{512} \approx 0.867446$ 
(the intercept for the dominant curve being simply 1), while the estimate of the true intercept using the numerically-generated curve is 
0.612243, quite close to our previously conjectured value of 
$\frac{135 \pi^2}{2176} \approx 0.612315$ \cite{slater833}.

In obtaining our several results, we used the "Bloore/correlation"  parameterization of density matrices 
\cite{bloore,joe,Rousseeuw} and accompanying ranges of integration--generated by the {\it cylindrical algebraic decomposition} procedure \cite{cylindrical,strzebonski}, implementing the requirement that $\rho$ be nonnegative definite--presented in \cite[eqs. (3)-(5)]{slaterPRA2}. The  computational tractability of utilizing the $3 \times 3$ principal minors of the partial transpose in this coordinate frame appeared to stem from the fact that each of these four quantities only contains  
{\it three} of the six off-diagonal variables ($z_{ij}$) employed in the full parameterization (each set of three variables, additionally and conveniently, sharing a common row/column subscript). (The nine-dimensional convex set of real two-qubit density matrices is parameterized by six off-diagonal--$z_{ij}=\frac{\rho_{ij}}{\sqrt{\rho_{ii} \rho_{jj}}}$--and three diagonal variables--$\rho_{ii}$.) Integrating out the three variables not present in the constraint simply leaves us with a constrained (boolean) integration over the cube $[-1,1]^3$, as indicated in \cite[eq. (3)]{slaterPRA2}--see (\ref{cube}) also. We appropriately permuted the subscripts in the indicated coordinate system, 
so that we could study all four of the minors (thus, finding that they fell into two equal sets). Of course, such a simplifying integration strategy is not available for the determinant of the partial transpose itself, which contains all the six off-diagonal variables ($z_{ij}$), rather than simply three.

Each of the constrained integrations we had initially used, employed as its constraint the nonnegativity of a {\it single} $2 \times 2$ or 
$3 \times 3$ principal minor of the partial transpose of $\rho$. (However, we were able above to 
splice together results, taking the 
sharper/tighter bounds over the half-axes provided by individual outcomes.) We had initially been unable--using either the (Bloore 
\cite{bloore}) density-matrix parameterization presented in 
\cite{slaterPRA2} or the interesting partial-correlation parameterization indicated in \cite{joe}--to perform constrained integrations in which two or more $2 \times 2$ or $3 \times 3$ minors (and {\it a fortiori} the determinant) are required to be {\it simultaneously} nonnegative. 
(It, then, remained an open question whether or not being able to do so would simply lead to the dominant and intermediate curves given already in Fig.~\ref{fig:threecurves} and by 
(\ref{dominantcurve}) and (\ref{intermediatecurve}). However, we were able eventually to answer this question positively for the $2 \times 2$ minors.)

We can, however, rather convincingly--but in a somewhat heuristic manner--reduce the derived upper bound on the HS separability probability of generic real two-qubit systems from $\frac{22}{35} \approx 0.628571$ to 0.576219 by using a new curve--having a $y$-intercept of $(\frac{45 \pi^2}{512})^2 \approx 0.752462$ as a DESF. (We apply a similar independence ansatz at the very end of the paper with quite interesting results [Fig.~\ref{fig:veryclose}].) This curve is obtained by taking the {\it product} of the two curves displayed in Fig.~\ref{fig:SymmetricCurves} (that is, the product of the function (\ref{intermediatecurve}) with its reflection about $\xi=0$). A plot of the result shows that
it is both subordinate to the intermediate curve in Fig.~\ref{fig:threecurves}, as is obvious it must be, but also clearly dominates the numerically-generated curve there, which is an estimate of the true DESF. (Since each of the two curves in Fig.~\ref{fig:SymmetricCurves2} is simply unity over a half-axis, a parallel strategy in the $2 \times 2$ minors analysis can, of course, yield no nontrivial upper-bound reduction from 
$\frac{1024}{135 \pi^2} \approx 0.76854$.)

The "twofold-ratio" theorem of Szarek, Bengtsson and 
{\.Z}yczkowski \cite{sbz}--motivated by the numerical results reported in \cite{slaterPRA}--allows us to immediately obtain exact upper bounds, as well, on the HS separability probability for generic (eight-dimensional) real 
{\it minimally-degenerate} real two-qubit systems (boundary states having a single eigenvalue zero). These upper bounds would, then, be {\it one-half} those applicable to the nondegenerate case--that is, $\frac{512}{135 \pi^2} \approx 0.38427$ and $\frac{11}{35} 
\approx 0.314286$. Further, we can, using the results of our numerical study, similarly obtain an induced estimate, 0.226421, of the true probability.

The two sets of derived functions (\ref{dominantcurve}) and (\ref{intermediatecurve}), based 
respectively on the $2 \times 2$ and $3 \times 3$ minors have the same functional forms, but with differing sets of constants 
($\{1,2,3,1\}$ {\it vs.} $\{9,2048=2^{11},27,7\}$). It seems natural, then, to conjecture that the true separability function--which must be based on the determinant of the partial transpose \cite[eq. (7)]{slaterPRA2} \cite{augusiak,ver}, that is, the single $4 \times 4$ minor--will also adhere to the same functional form, but with a different set of constants.

\begin{figure}
\includegraphics{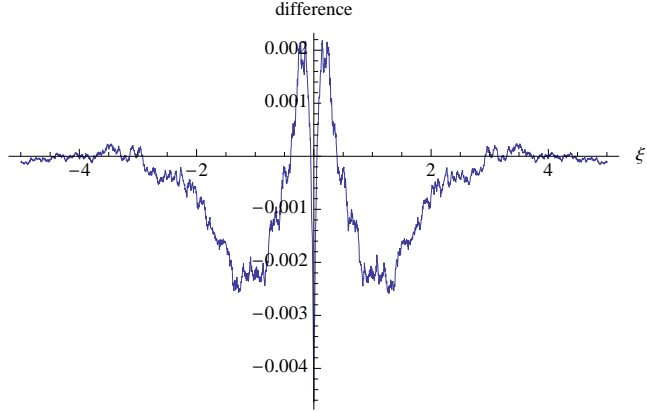}
\caption{\label{fig:differences}The difference between the numerically-generated subordinate function in Fig.~\ref{fig:differences} and a hypothetically true separability function (\ref{conjecture})--fitting a general pattern observed--giving a separability probability of $\frac{29}{64} \approx 0.453125$}
\end{figure}

\begin{figure}
\includegraphics{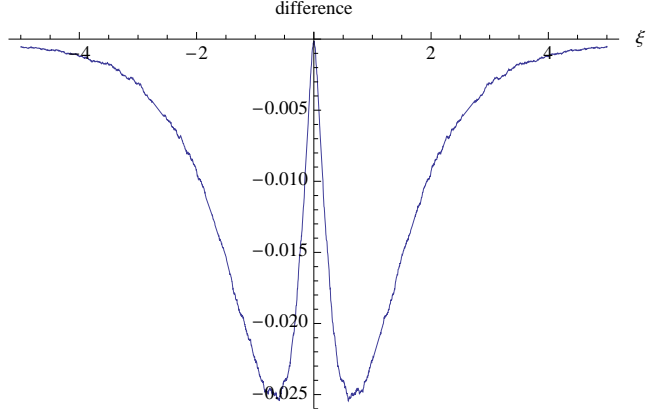}
\caption{\label{fig:differences2}The difference between the numerically-generated subordinate function in Fig.~\ref{fig:differences} and the previously conjectured true separability function (\ref{conjecture2}), giving a separability probability of $\frac{8}{17} \approx 0.470588$, and a poorer fit than Fig.~\ref{fig:differences}}
\end{figure}

In fact, pursuing this line of thought, as an exercise, we have found that the function 
\begin{equation} \label{conjecture}
S_{conjecture}(\xi)=\begin{cases}
\frac{315 e^{-3 \xi } \left(-5+18 e^{2 \xi }\right) \pi ^2}{2^{16}} & \xi
   >0 \\
 -\frac{315 e^{\xi } \left(-18+5 e^{2 \xi }\right) \pi ^2}{2^{16}} & \xi
   <0 
 \end{cases}
\end{equation}
fits (Fig.~\ref{fig:differences}) the numerically-generated subordinate curve in Fig.~\ref{fig:threecurves} quite well, yielding an HS separability probability of 
$\frac{29}{64} \equiv \frac{29}{2^6} \approx  0.453125$, and a $y$-intercept of $\frac{4095 \pi^2}{2^{16}} \approx 0.6167$.  (Then, by the twofold-ratio theorem \cite{sbz}, the HS separability probability of the minimally-degenerate (boundary) states would be $\frac{29}{128} \equiv \frac{29}{2^7}  \approx 0.226563$. Also, we have been able to find a number of other curves, adhering to this same general structure, fitting the subordinate curve in Fig.~\ref{fig:threecurves} equally as well, and again  yielding $\frac{29}{64}$ as a separability probability, in addition to 
well-fitting curves yielding somewhat less simple fractions--such 
as $\frac{163}{360} \approx 0.452778, \frac{367}{810} \approx 0.453086$ and
$\frac{428}{945} \approx 0.45291$.)
We are obligated, however, to note that in 
\cite[sec. IX.A]{slater833} we had advanced--based on somewhat different considerations (scaling constants, in particular) than here--the hypothesis that this probability is 
$\frac{8}{17} \approx 0.470588$, with an associated DESF equal to 
\begin{equation} \label{conjecture2}
S_{previous}(\xi)=\begin{cases}
\frac{135 e^{-3 \xi } \left(-1+3 e^{2 \xi }\right) \pi ^2}{2^8 \cdot 17} & \xi
   >0 \\
 -\frac{135 e^{\xi } \left(-3+e^{2 \xi }\right) \pi ^2}{2^8 \cdot 17} & \xi <0 
\end{cases}.
\end{equation}
(However, our best numerical estimate at that point was 0.4538838 \cite[sec. V.A.2]{slaterPRA2} \cite[sec. IX.A]{slater833}, rather close to our current-study estimate of 0.4528427. By computing standard errors of the mean, we can establish a [$\approx 95\%$] confidence range--four standard deviations wide--for this latter estimate of $(0.451634, 0.454051)$--that does contain 
$\frac{29}{64} \approx 0.453125$.
A comparable plot (Fig.~\ref{fig:differences2}) to 
Fig.~\ref{fig:differences} shows (\ref{conjecture2}) to provide a considerably poorer fit.)

One might further speculate--in line with random matrix theory and our previous analyses \cite{slater833}--that the DESF for the generic (15-dimensional) {\it complex} two-qubit systems is proportional to the {\it square} of (\ref{conjecture}). If the constant of proportionality were simply taken to equal unity, the associated HS separability probability, using the measure (\ref{andaiformula}) with $\beta=2$, would be
$\frac{30660525 \pi ^4}{11811160064}= \frac{3^5 \cdot 5^2 \cdot 7^2 \cdot 103 \pi ^4}{ 2^{30} \cdot 11} \approx 0.252864$, rather close to the value 
$\frac{8}{33} \approx 0.242424$ conjectured, for a number of reasons, in \cite[sec. IX.B]{slater833}. Proceeding similarly, using the fourth power of (\ref{conjecture}), rather than the square, and the measure (\ref{andaiformula}) with $\beta=4$, we obtain the HS {\it quaternionic} probability analogue of $\frac{4893927891755175 \pi ^8}{535315866107766636544} =\frac{3^{10} \cdot 5^2 \cdot 7^5 \cdot 13 \cdot 15173 \pi^8}{2^{56} \cdot 17 \cdot 19 \cdot 23} \approx 0.0867454$.

Duplicating the line of analysis of the immediately preceding paragraph, but now using the intermediate curve (\ref{intermediatecurve}) given in Fig.~\ref{fig:threecurves}, instead of the conjectured curve 
(\ref{conjecture})--and taking the constant of proportionality again to equal 1--we obtain tentative (induced) exact upper bounds on the HS separability probability for the complex two-qubit states of
 $\frac{752517 \pi ^4}{149946368} \approx 0.488855$ and 
$\frac{14092854769917 \pi ^8}{408413594137395200} \approx 0.327414$ for the quaternionic two-qubit states.

In \cite{slater833}, we studied several two-qubit real, complex and mixed scenarios, in which--in order to obtain exact HS separability probabilities--certain of the off-diagonal entries were {\it a priori} set to zero. In one such (7-dimensional) scenario, we nullified  four of the off-diagonal entries, allowing only the (1,4)- and (2,3)-entries (the ones interchanged under partial transposition) to be complex 
\cite[sec. II.B.3]{slater833}. The associated HS separability probability was 
$\frac{2}{5}$. We have now been able--parameterizing the off-diagonal entries using polar coordinates--to extend this 7-dimensional scenario to a 9-dimensional one, allowing, additionally, any single one of the remaining  four off-diagonal entries ((1,2), (1,3), (2,4) or (3,4)) to be arbitrary complex. The associated DESF is 
\begin{equation}
S(\xi)= \begin{cases}
 \frac{1}{3} e^{-4 \xi } \left(-1+4 e^{2 \xi }\right) & \xi >0 \\
 -\frac{1}{3} e^{2 \xi } \left(-4+e^{2 \xi }\right) & \xi <0
\end{cases},
\end{equation}
with an accompanying 
HS separability probability of 
$\frac{17}{35} \approx 0.485714$.

\begin{figure}
\includegraphics{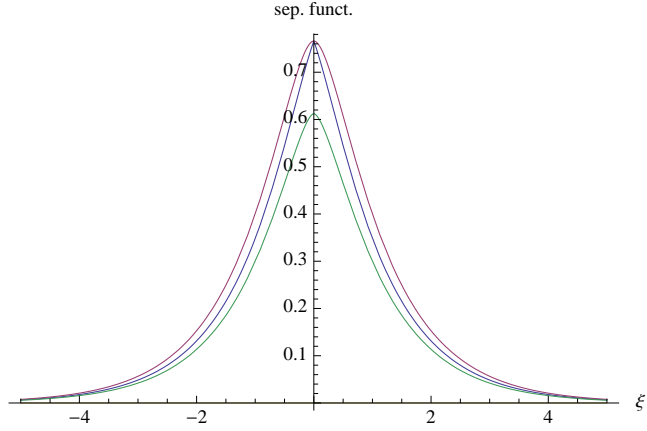}
\caption{\label{fig:pairedminors}The subordinate curve is the same numerically-derived DESF estimate of the true separability probability displayed as the subordinate curve in Fig.~\ref{fig:threecurves}. The other two curves are obtained by 
{\it simultaneously} enforcing the nonnegativity of certain pairs of the four $3 \times 3$ principal minors of the partial transpose of $\rho$. 
Both of these superior curves intercept the $\xi$-axis at $\frac{11127 \pi ^2}{143360} \approx 0.766037$.}
\end{figure}

Let us now present an additional figure (Fig.~\ref{fig:pairedminors}) showing--as in 
Fig.~\ref{fig:threecurves}--three DESF's. 
The subordinate curve in this new figure is identically the same as the subordinate numerically estimated curve in Fig.~\ref{fig:threecurves}. The dominant of the three curves 
 has the form
\begin{equation} \label{newdominant}
\begin{cases}
 -\frac{\pi  e^{-6 \xi } \left(\sqrt{e^{2 \xi }-1} \left(1696 e^{2 \xi
   }-7665 e^{4 \xi }-5346 e^{6 \xi }+188\right)+3 e^{4 \xi } \left(-7273
   e^{2 \xi }+1782 e^{4 \xi }+1782\right) \csc ^{-1}\left(e^{\xi
   }\right)\right)}{71680} & \xi >0 \\
 -\frac{\pi  e^{-2 \xi } \left(e^{\xi } \sqrt{1-e^{2 \xi }} \left(-7665
   e^{2 \xi }+1696 e^{4 \xi }+188 e^{6 \xi }-5346\right)+3 \left(-7273
   e^{2 \xi }+1782 e^{4 \xi }+1782\right) \sin ^{-1}\left(e^{\xi
   }\right)\right)}{71680} & \xi <0
\end{cases}
\end{equation}
and the intermediate curve (obtained, as earlier [cf. Fig.~\ref{fig:SymmetricCurves}] by splicing together the
 $\xi<0$ and $\xi>0$ {\it lesser} branches of two curves equal under reflection around $\xi=0$), the form
\begin{equation} \label{newintermediate}
\begin{cases}
 \frac{3 \pi ^2 e^{-3 \xi } \left(18873 e^{2 \xi }-4037\right)}{573440}
   & \xi >0 \\
 \frac{3 \pi ^2 e^{\xi } \left(18873-4037 e^{2 \xi }\right)}{573440} &
   \xi <0
\end{cases}.
\end{equation}
(This does possess the same basic functional form as already encountered in (\ref{dominantcurve}), (\ref{intermediatecurve}) and (\ref{conjecture})).
Again, as in Fig.~\ref{fig:threecurves}, of course, the numerically-generated curve yields a separability probability estimate of 0.4528427, while the dominant  curve yields 0.585542, and the intermediate curve, an exact value of 
$\frac{1129}{2100} = \frac{1129}{2^2 \cdot 3  \cdot 5^2 \cdot 7} \approx 0.537619$. (Both of these last two curves intercept the $\xi$-axis at $\frac{11127 \pi ^2}{143360} \approx 0.766037$.) Thus, $\frac{1129}{2100} <
\frac{22}{35}$ provides a further improved exact upper bound on the true Hilbert-Schmidt separability probability.
The two superior curves in Fig.~\ref{fig:pairedminors} are obtained by enforcing {\it simultaneously} the nonnegativity of certain pairs of the four $3 \times 3$ principal minors of the partial transpose of $\rho$.  The dominant curve (\ref{newdominant}) is derived by pairing the first and second minors (or, equivalently, three other possible pairs), while the intermediate  curve (\ref{newintermediate}) is achieved uniquely by coupling (taking the lesser branches) the results pairing the second with the third minor with the (reflection-symmetric) 
results pairing the first with the fourth minor. (By the $k$-th minor we mean the one obtained by elimination from the partial transpose of $\rho$ of its $k$-th row and column.)

Given that the direct/naive enforcement of the simultaneous nonnegativity of pairs of $3 \times 3$ principal minors (requiring, 
a constrained {\it five}-dimensional integration) appeared to be intractable, we resorted to an alternative strategy to obtain the two superior curves in Fig.~\ref{fig:pairedminors}.
We exploited the fact already noted above that each of the four $3 \times 3$ principal minors is parameterized by only three (of the six) off-diagonal Bloore (correlation) variables $z_{ij}$'s, with all of the three sharing a common row/column index 
(such as the common $i$ index in $z_{ij}, z_{ik}, z_{il}$ {\it etc.}), for example, the 4-th minor (with $i=1,j=2,k=3,l=4$) takes the form
\begin{equation} \label{sampleminor}
\mbox{minor}_{3 \times 3} = 2 e^{\xi } z_{ij} z_{ik} z_{il}-z_{ij}^2-z_{ik}^2-e^{2 \xi } z_{il}^2+1.
\end{equation}
We can, then,  arrange--by using a suitably chosen cylindrical algebraic decomposition--that any such set of three variables (sharing a common index) comprises the last three to be integrated over of the six variables. By performing the first (unconstrained) three of the six one-dimensional integrations (over, say, $z_{jk}, z_{jl}$ and $z_{kl}$, in our example), we are simply left with 
(cubical) integrations  of the form
\begin{equation} \label{cube}
\int_{-1}^{1} \int_{-1}^{1} \int_{-1}^{1}  (\frac{3}{4})^3 (1-z_{ij}^2) (1-z_{ik}^2) (1-z_{il}^2) dz_{ij} dz_{ik} dz_{il} =1.
\end{equation}
(By reparameterizing the $z_{ij}$'s in terms of {\it partial} correlations \cite{joe}, one could re-express the 
full six-dimensional integration as the integral of a simple product measure over a {\it six}-dimensional hypercube. But, then,
the nonnegativity requirements on the partial transpose appear to take on quite cumbersome forms.)
Only, at this stage of integration--after having integrated out three (extraneous) variables--do we then need to impose (inside the integral signs) the three-dimensional nonnegativity requirement of a {\it single} minor 
($\mbox{minor}_{3 \times 3} \geq 0$), such as (\ref{sampleminor}) to obtain the results reported earlier here.

To further proceed, in our scheme,
we perform the outer two (over $z_{il}$ and $z_{ik}$, in our example) 
of the three indicated integration steps in (\ref{cube})--and its analogues--over the corresponding cubes for each of  two paired minors {\it independently} of one another, and then combine (multiply) the two results together, which are then integrated (in a joint manner) over the remaining {\it shared}/last variable ($z_{ij}$ in our illustration here) to derive the new curves in Fig.~\ref{fig:pairedminors}. Our approach here, thus,  consists in replacing a direct [but intractable] five-dimensional constrained integration--five being the number of variables parameterizing any two of the four $3 \times 3$ principal minors, we want to be simultaneously nonnegative--by a pair of 
independent constrained two-dimensional integrations (each member of the pair concerned with the nonnegativity of only a single minor) conducted over three-dimensional cubes. The two distinct one-dimensional results obtained are, then, joined by multiplication together  into a single one-dimensional integration (over $z_{ij}$, the shared variable, in our illustration). Since there is a factor of $(\frac{3}{4})^3=\frac{27}{64}$ in the three-fold integrals (\ref{cube}), we importantly assign--by symmetry--a weight of $\frac{3}{4}$ to each single-fold integration step taken.

\begin{figure}
\includegraphics{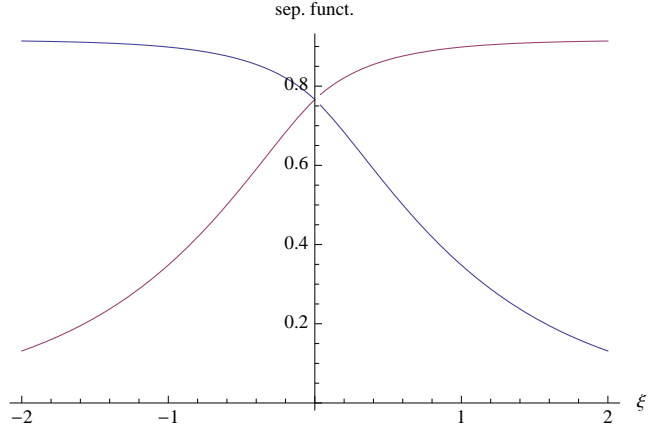}
\caption{\label{fig:pairedminors2}The two distinct (red and blue) separability functions obtained from enforcing the joint nonnegativity of the first and fourth $3 \times 3$ principal minors of the PT--giving the dominant (blue) curve on the left--as well as the joint nonnegativity of the second and third $3 \times 3$ principal minors of the PT. The lesser branches of the two curves define the intermediate curve (\ref{newintermediate}) in Fig.~\ref{fig:pairedminors}. The greater branches are described by (\ref{greaterbranches}). The value at the intersection is $\frac{11127 \pi ^2}{143360} \approx 0.766037$.}
\end{figure}

\begin{figure}
\includegraphics{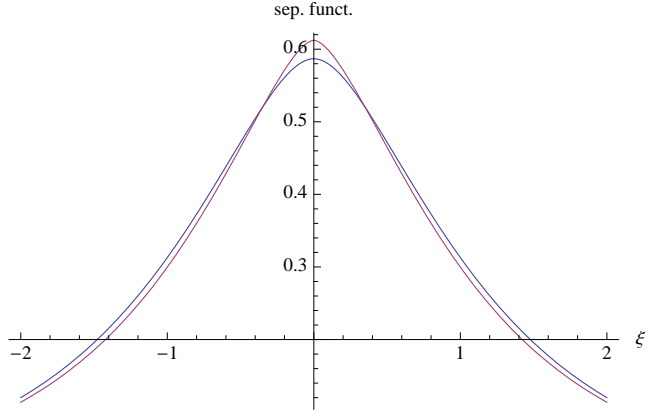}
\caption{\label{fig:veryclose}The (blue) separability function derived by taking the product of the two curves displayed in Fig.~\ref{fig:pairedminors2} along with the numerical estimate of the true separability function. The latter (red) curve crosses the $y$-intercept at 0.612243 and the (blue) product
DESF at the lesser value of $\frac{123810129 \pi ^4}{2^{24} \cdot 5^2 \cdot 7^2} \approx 0.586813$.}
\end{figure}

The intermediate curve in Fig.~\ref{fig:pairedminors}, given by (\ref{newintermediate}), is constructed by
taking the lesser branches of the two curves in Fig.~\ref{fig:pairedminors2}. (The four possible pairings of minors other than the first with the fourth, and the second with the third all yield the same dominant curve shown in 
Fig.~\ref{fig:pairedminors}.)
The two {\it greater} branches in Fig.~\ref{fig:pairedminors2}, together 
yielding an upper bound on
the separability probability of $\frac{7724}{525}-\frac{5751 \pi ^2}{4096} \approx 0.854936$, take the form
\begin{equation} \label{greaterbranches}
\begin{cases}
 -\frac{3 \pi ^2 e^{-6 \xi } \left(3 e^{2 \xi } \left(91 e^{2 \xi }
   \left(9-65 e^{2 \xi }\right)+144\right)+20\right)}{573440} & \xi >0 \\
 -\frac{3 \pi ^2 \left(2457 e^{2 \xi }+432 e^{4 \xi }+20 e^{6 \xi
   }-17745\right)}{573440} & \xi <0
\end{cases}.
\end{equation}

Let us--similarly as we have done before--take as a new separability function the {\it product}
of the two curves displayed in Fig.~\ref{fig:pairedminors2}. Use of this product DESF in formula 
(\ref{JACOBIAN}) yields
\begin{equation}
P^{HS}_{sep/real}=\frac{\pi ^2 \left(18031791 \pi ^2-177044420\right)}{2^{14} \cdot 5^2 \cdot 7^2} \approx 
0.453503,
\end{equation}
{\it very} close to our earlier numerical estimates of 
0.4538838 \cite[sec. V.A.2]{slaterPRA2} \cite[sec. IX.A]{slater833} and lying {\it within}
the confidence range  $(0.451634, 0.454051)$, we established above. (Possibly, this product DESF is, in fact, the function that would arise if one could {\it simultaneously} enforce the nonnegativity of {\it all} four $3 \times 3$ principal minors [but see final paragraph]. However, it lacks the simple functional form repeatedly previously observed above.)
We display this derived product separability function in Fig.~\ref{fig:veryclose} along with the closely-fitting numerical estimate of the true function, already appearing as the subordinate function in both 
Figs.~\ref{fig:threecurves} and \ref{fig:pairedminors}. (If we, in a similar vein, take as a product DESF, the {\it square} of (\ref{newdominant}), that is, the dominant curve in Fig.~\ref{fig:pairedminors}--since this arises identically from four of the six possible pairings of minors--the associated separability probability falls, rather unrealistically to 0.367762. So, {\it independence} of minor pairings does not appear to be a tenable hypothesis in this case.)

It does, however, appear that we can reduce the $y$-intercept in 
Fig.~\ref{fig:pairedminors} from $\frac{11127 \pi ^2}{143360} \approx 0.766037$ to $\frac{159104}{231525} \approx 0.6872$ by enforcing the simultaneous nonnegativity of the second, third and fourth minors,  using repeated integration over cubes.

\begin{acknowledgments}
I would like to express appreciation to the Kavli Institute for Theoretical
Physics (KITP)
for computational support in this research.
\end{acknowledgments}

\bibliography{Minors6}

\end{document}